\documentclass[twocolumn,showpacs,showkeys,amsmath,amssymb,pra]{revtex4-1}
      
\usepackage{color}
\usepackage{bm}
\usepackage{graphicx}   	

\newcommand{\rd}{{\mathrm d}}

\newcommand{\rt}{(\bm r;t)} 
\newcommand{\ph}{\widehat{\psi}}
\newcommand{\phd}{\widehat{\psi}^\dagger}

\newcommand{\Uh}{\widehat{U}}
\newcommand{\vac}{| {\rm vac} \rangle}
\newcommand{\wP}{\widetilde{\Psi}}
\newcommand{\aL}{a_1^{\phantom{\dagger}}}
\newcommand{\aR}{a_2^{\phantom{\dagger}}}
\newcommand{\aj}{a_j^{\phantom{\dagger}}}
\newcommand{\aLd}{a_1^\dagger}
\newcommand{\aRd}{a_2^\dagger}
\newcommand{\ajd}{a_j^\dagger}              
\begin{document}

\title[Macroscopic wave functions]
	{Emergence and destruction of macroscopic wave functions}

\author{Bettina Gertjerenken}	
\author{Martin Holthaus}
%\email[e-mail: ]{martin.holthaus@uni-oldenburg.de}

\affiliation{Institut f\"ur Physik, 
	Carl von Ossietzky Universit\"at, 
	D-26111 Oldenburg, Germany}

\date{July 27, 2015}
		    
\begin{abstract}
The concept of the macroscopic wave function is a key for understanding 
macroscopic quantum phenomena. The existence of this object reflects a 
certain order, as is present in a Bose-Einstein condensate when a 
single-particle orbital is occupied by a macroscopic number of bosons. 
We extend these ideas to situations in which a condensate is acted on by an 
explicitly time-dependent force. While one might assume that such a force 
would necessarily degrade any pre-existing order, we demonstrate that 
macroscopic wave functions can persist even under strong forcing. Our 
definition of the time-dependent order parameter is based on a comparison 
of the evolution of $N$-particle states on the one hand, and of states with 
$N - 1$ particles on the other. Our simulations predict the possibility of 
an almost instantaneous dynamical destruction of a macroscopic wave function 
under currently accessible experimental conditions.
\end{abstract} 


% 03.75.Kk 	Dynamic properties of condensates; collective and hydrodynamic 
%		excitations, superfluid flow
% 03.75.Lm 	Tunneling, Josephson effect, Bose-Einstein condensates in 
%               periodic potentials, solitons, vortices, and topological 
%               excitations
% 67.85.De 	Dynamic properties of condensates; excitations, and superfluid 
%               flow


\maketitle 

%%%%%%%%%%%%%%%%%%%%%%%%%%%%%%%%%%%%%%%%%%%%%%%%%%%%%%%%%%%%%%%%%%%%%%%%%%%%%%%%

%%%%%%%%%%%%%%%%%%%%%%%%%%%
%% INTRODUCTION
%%%%%%%%%%%%%%%%%%%%%%%%%%%

\section{Introduction}
Superconductors, superfluids, and atomic Bose-Einstein condensates are 
described in terms of a macroscopic wave function, a notion originally 
conceived in  London's theory of superfluidity~\cite{London64}: Instead of 
considering the Schr\"odinger wave function $\Psi(\bm r_1, \ldots, \bm r_N; t)$
of a Bose-condensed interacting $N$-particle system, one works with an 
effective single-particle wave function $\Phi\rt$ which obeys the nonlinear 
Gross-Pitaevskii equation~\cite{Pitaevskii61,Gross61,Gross63,PethickSmith08,
PitaevskiiStringari03}. Experimental justification for the concept of the 
macroscopic wave function is provided by the observation of Josephson 
tunneling~\cite{Josephson62} between two superconductors coupled by a 
weak link~\cite{AndersonRowell63}. Moreover, the occurrence of vortices, 
as observed in a series of landmark experiments with Bose-Einstein 
condensates~\cite{MatthewsEtAl99,MadisonEtAl00,AboShaeerEtAl01}, is a direct 
consequence of the existence of a macroscopic wave function. Obviously, the 
reduction of the full $N$-particle dynamics to that of a single-particle 
wave function requires that the system under consideration is highly 
{\em ordered\/}. This order is connected to the idea that $\Phi\rt$ represents 
a macroscopically occupied single-particle orbital, so that the terms 
``macroscopic wave function'' and ``order parameter'' often are used 
synonymously~\cite{Leggett00}.
 
But now new experimental developments are posing new theoretical challenges.
There is an increasing tendency to subject Bose-Einstein condensates to 
strong time-dependent forcing, so as to ``engineer'' novel systems which may 
not be accessible without such forcing. For instance, dynamic localization 
and quasienergy band engineering has been demonstrated with Bose-Einstein 
condensates in strongly shaken optical lattices~\cite{LignierEtAl07,
EckardtEtAl09}, and coherent control over the superfluid-to-Mott insulator
transition has been achieved~\cite{EckardtEtAl05,ZenesiniEtAl09}. Moreover,
giant Bloch oscillations have been realized with condensates in tilted, 
ac-driven optical lattices~\cite{AlbertiEtAl09,HallerEtAl10}. Still further 
experiments have demonstrated time-reversal symmetry breaking in shaken 
triangular lattices~\cite{StruckEtAl11}, and controlled photon-assisted 
tunneling~\cite{MaEtAl11,ChenEtAl11}. A particularly ambitious line of this 
research addresses the realization and usage of tunable artifical gauge 
fields~\cite{StruckEtAl12,HaukeEtAl12,StruckEtAl13,GoldmanDalibard14} or, 
phrased more generally, the exploitation of Bose-Einstein condensates in 
strongly forced optical lattices for quantum simulation 
purposes~\cite{ParkerEtAl13}.

These activities lead to an important question: To what extent is the 
underlying order degraded if one subjects a macroscopic wave function to 
strong forcing? It has been emphasized already quite early that the solution 
to the time-dependent Gross-Pitaevskii equation does not represent a 
condensate if it becomes {\em chaotic\/}~\cite{CastinDum97}. 
The obvious conflict between dynamical chaos and the possible existence of an 
order parameter has inspired further studies especially on $\delta$-kicked 
condensates, both theoretical and experimental ones~\cite{ZhangEtAl04,
DuffyEtAl04,WimbergerEtAl05,ShresthaEtAl13}. But while the nonlinear 
Gross-Pitaevskii equation naturally can produce chaotic solutions in the 
presence of external forcing, the actual $N$-particle system still is 
described by a linear Schr\"odinger equation. Hence, while the $N$-particle 
wave functions cannot become chaotic in the sense of nonlinear dynamics, 
there should nonetheless be a certain quality of the time-dependent
$N$-particle system which decides whether or not the solution to the 
Gross-Pitaevskii equation actually qualifies as a macroscopic wave function, 
and there should be a measure which quantifies the degree of order remaining 
in a Bose-Einstein condensate under the action of an external force. In this 
letter we suggest an approach to these issues which does not involve the
familiar partitioning of the field operator into a condensate part and a 
noncondensate part~\cite{Gardiner97,CastinDum98,Leggett01,GardinerMorgan07,
BillamGardiner12,BillamEtAl13}, but focuses on the evolution of neighboring 
states in Fock space. This may be seen as similar in spirit to the 
characterization of the degree of chaos in classical dynamical systems
by probing the way initially close trajectories separate in time.

%%%%%%%%%%%%%%%%%%%%%%%%%%%
%% MAIN FINDINGS
%%%%%%%%%%%%%%%%%%%%%%%%%%%

\begin{figure}[t]
\includegraphics[width = 0.9\linewidth]{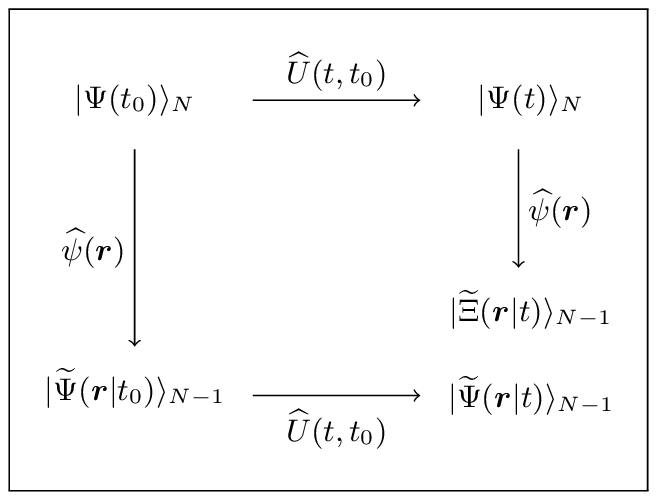}
\caption{{\textbf{Scheme for constructing the time-dependent macroscopic 
	wave function.}}
        An initial $N$-boson state $|\Psi(t_0)\rangle_N$ develops in time 
	according to the time-evolution operator $\Uh(t,t_0)$, giving 
	$|\Psi(t)\rangle_N$. If one acts with the field operator $\ph(\bm r)$ 
	on the initial state and normalizes, one obtains subsidiary 
	$(N-1)$-particle states $|\wP (\bm r | t_0) \rangle_{N-1}$, which 
	also propagate in time. A candidate function $\Phi\rt$ then is 
	introduced by taking the matrix elements of the field operator with 
	$|\Psi(t)\rangle_N$ and $|\wP (\bm r | t) \rangle_{N-1}$. 
	On the other hand, propagating first and annihilating thereafter 
	yields $|\widetilde{\Xi}(\bm r | t) \rangle_{N-1}$. If the absolute 
	value of the projection ${_{N-1}\langle} \wP (\bm r | t) | 
	\widetilde{\Xi}(\bm r | t) \rangle_{N-1}$ equals unity to good
	accuracy, $\Phi\rt$ actually is a macroscopic wave function which 
	obeys the Gross-Pitaevskii equation.}
\label{F_1}	
\end{figure}

\section{The order parameter}
A key question is how the time-dependent macroscopic wave function, 
if it exists, is obtained from the full $N$-particle state. A guide 
to the answer can be inferred from the discussion given by Lifshitz and 
Pitaevskii~\cite{LaLifIX}: Let $| \Psi(t) \rangle_N$ be a time-dependent 
$N$-particle condensate state, and let $| \wP(t) \rangle_{N-1}$ be a ``like'' 
state of $N-1$ particles; then the macroscopic wave function, normalized to 
unity, should be given by
\begin{equation} 
	\Phi\rt = \lim_{N \to \infty}
	{_{N-1}\langle} \wP(t) | \ph(\bm r) | \Psi(t)\rangle_N / \sqrt{N}
	\; ,
\label{eq:LPS}	
\end{equation}
where $\ph(\bm r)$ is the bosonic field operator. But this leaves open the 
question how to quantify the ``likeness'' of $| \Psi(t) \rangle_N$ and 
$| \wP(t) \rangle_{N-1}$; if two such states are ``like'' at one particular 
moment $t_0$, they might not remain so under the influence of time-dependent 
forcing. Moreover, it seems desirable to get rid of the limit of an infinite 
particle number, and to study the emergence of a ``macroscopic'' wave function 
already when~$N$ is relatively small. With this background, we proceed as 
summarized by Fig.~\ref{F_1}: We start from an initially given $N$-boson 
state $|\Psi(t_0)\rangle_N$, which is not necessarily a pure condensate.  
Under the influence of some force which does not need to be specified at this 
point it develops in time into the $N$-particle state $|\Psi(t)\rangle_N$, 
as determined by the system's time-evolution operator $\Uh(t,t_0)$. 
In order to generate suitable $(N-1)$-particle states for taking
the matrix elements suggested by Eq.~(\ref{eq:LPS}), we act with the bosonic 
annihilation operators $\ph(\bm r)$ on the initial state, and normalize the 
results, obtaining
\begin{equation}
	| \wP(\bm r | t_0) \rangle_{N-1} = 	
	\frac{\ph(\bm r) | \Psi(t_0)\rangle_N}
	     {\| \ph(\bm r) | \Psi(t_0)\rangle_N \|} \; . 
\label{eq:SUB}
\end{equation}
These subsidiary states likewise evolve in time under the action of the very 
same evolution operator, giving states $| \wP(\bm r | t) \rangle_{N-1}$. We 
then define a function $\Phi\rt$ according to
\begin{equation}		
	\sqrt{N} \Phi\rt = 
	{_{N-1}\langle} \wP (\bm r|t) | \ph(\bm r) | \Psi(t) \rangle_{N}	
	\; .
\label{eq:DEF}
\end{equation}
Observe the difference to the above tentative prescription~(\ref{eq:LPS}): We 
employ not just one single subsidiary $(N-1)$-particle state, but infinitely 
many; in principle, there is one state $| \wP(\bm r | t) \rangle_{N-1}$ 
associated with each $\bm r$ considered. Still, $\Phi\rt$ as defined by 
Eq.~(\ref{eq:DEF}) should qualify as a macroscopic wave  function for 
sufficiently large~$N$, and obey the Gross-Pitaevskii equation, 
{\em provided\/} the above ``likeness''-condition is satisfied. This means 
that the state trajectories evolving from the respective initial states 
$| \wP(\bm r | t_0) \rangle_{N-1}$ and $| \Psi(t_0) \rangle_{N}$ in Fock 
space should not diverge from each other too much, in a suitable sense. 
To bring this intuitive idea into a precise form, we also annihilate a boson 
from the time-evolved $N$-particle state, thus producing 
\begin{eqnarray}
 	| \widetilde\Xi(\bm r | t) \rangle_{N-1} & = &
	\frac{\ph(\bm r) | \Psi(t) \rangle_N}
	     {\| \ph(\bm r) | \Psi(t) \rangle_N \|}
\nonumber \\	& = &
	\frac{\ph(\bm r) \Uh(t,t_0) | \Psi(t_0) \rangle_N}
	     {\| \ph(\bm r) | \Psi(t) \rangle_N \|} \; . 		     
\end{eqnarray}
Then the scalar products
\begin{eqnarray}
\label{eq:TDO}
	R\rt & = & {_{N-1}\langle} \wP (\bm r | t) |
	\widetilde{\Xi}(\bm r | t) \rangle_{N-1} 
	\phantom{\sum}
\\		& = &
	\frac{{_N\langle} \Psi(t_0) | \phd(\bm r) \widehat{U}^\dagger(t,t_0) 
	        \ph(\bm r) \Uh(t,t_0) | \Psi(t_0) \rangle_N}
	     {\| \ph(\bm r) | \Psi(t_0) \rangle_N \| \; 
	      \| \ph(\bm r) | \Psi(t) \rangle_N \| } 
\nonumber	      	     	
\end{eqnarray}
have a particular significance: If $|R\rt| = 1$, the candidate $\Phi\rt$
provided by Eq.~(\ref{eq:DEF}) is a true macroscopic wave function, 
obeying the Gross-Pitaevskii equation. In general, the magnitude $|R\rt|$, 
varying between $0$ and $1$, provides the desired measure of the degree of 
order of the time-evolving $N$-boson system.

The justification for this statement stems from the observation that the
proper macroscopic wave function has to satisfy the requirement    
\begin{equation}
	N | \Phi\rt |^2 = 
	{_N\langle} \Psi(t) | \phd(\bm r) \ph(\bm r) | \Psi(t) \rangle_N \; , 
\label{eq:REQ}
\end{equation}
demanding that its absolute square, multiplied by the particle number~$N$,
yields the exact $N$-particle density of the system~\cite{LaLifIX}. 
Introducing the projection operators  
\begin{equation}
	\widehat{Q}_t = 
	| \widetilde\Xi(\bm r| t) \rangle_{N-1} \; 
	{_{N-1}\langle} \widetilde\Xi(\bm r| t)| \; ,
\label{eq:POQ}
\end{equation}
we have the obvious identity 
\begin{eqnarray} & &
	{_N\langle} \Psi(t) | \phd(\bm r) \ph(\bm r) | \Psi(t) \rangle_N 
\nonumber \\	& = &
	{_N\langle} \Psi(t) | \phd(\bm r) \widehat{Q}_t \ph(\bm r) | 
	\Psi(t) \rangle_N .
\end{eqnarray}	
Now, if this projector~(\ref{eq:POQ}) were equal to the projector 
$\widehat{P}_t$ defined by  
\begin{equation}
	\widehat{P}_t =
	| \wP(\bm r | t) \rangle_{N-1} \; {_{N-1}\langle} \wP(\bm r | t) | \; ,
\end{equation}	
which, in its turn, would be the case if 
$| \widetilde\Xi(\bm r| t) \rangle_{N-1}$ differed from 
$| \wP(\bm r | t) \rangle_{N-1}$ 
by not more than a phase factor, we could deduce
\begin{eqnarray} & &
	{_N\langle} \Psi(t) | \phd(\bm r) \ph(\bm r) | \Psi(t) \rangle_N 
\nonumber \\	& = & 
	{_N\langle} \Psi(t) | \phd(\bm r) \widehat{P}_t \ph(\bm r) | 
	\Psi(t) \rangle_N , 
\end{eqnarray}
from which the desired identity~(\ref{eq:REQ}) follows immediately, keeping
in mind the definition~(\ref{eq:DEF}). 

This reasoning deserves still more scrutiny. Namely, {\em if\/}   
$| \widetilde\Xi(\bm r| t) \rangle_{N-1}$ indeed differs from 
$| \wP(\bm r | t) \rangle_{N-1}$ merely by a phase factor, then 
$\ph(\bm r) | \Psi(t) \rangle_N$ is proportional to 
$| \wP(\bm r | t) \rangle_{N-1}$, wherefrom one is led to the relation 
\begin{equation}
	\ph(\bm r) | \Psi(t) \rangle_N = 
	\sqrt{N}\Phi\rt | \wP(\bm r | t) \rangle_{N-1} \; .
\label{eq:TCO}
\end{equation} 
This is reminiscent of what defines a condensate: Assuming that the 
$N$-particle state at some moment $t_0$ corresponds to an $N$-fold occupied 
single-particle orbital $\varphi(\bm r, t_0)$ and thus is a pure condensate 
state of the form
\begin{equation}
	| \Psi_\varphi (t_0) \rangle_{N} = \frac{1}{\sqrt{N!}}
	\left[ 
	\int \! \rd^3 r \, \varphi(\bm r; t_0) \, \phd(\bm r) 
	\right]^N \vac \; , 
\label{eq:NPC}	
\end{equation}
it obeys the equation 
\begin{equation}
	\ph(\bm r) | \Psi_\varphi (t_0) \rangle_{N} = 
	\sqrt{N} \varphi(\bm r;t_0) | \Psi_\varphi (t_0) \rangle_{N-1}   	
\label{eq:NCO}
\end{equation}
at that moment $t_0$. Therefore, the condition~(\ref{eq:TCO}), which (if 
satisfied) guarantees that the time-dependent order parameter takes on its 
maximum value $|R\rt| = 1$, and thus makes sure that the candidate $\Phi\rt$ 
defined through Eq.~(\ref{eq:DEF}) actually is a macroscopic wave function, 
generalizes the familiar characterization of a pure condensate expressed by 
Eq.~(\ref{eq:NCO}) so as to also involve time evolution, and reduces to it 
when the time~$t$ is close to $t_0$. Again adopting the dynamical-systems 
viewpoint, the projection~(\ref{eq:TDO}) compares the trajectory of the 
given $N$-particle state in Fock space to that of subsidiary, neighboring 
$(N-1)$-particle states. If it does not matter whether one annihilates first 
and propagates then, or whether one propagates prior to annihilating, the flow 
in Fock space may be considered as (locally) stiff. Hence, we refer to the 
magnitude $|R\rt|$ as {\em stiffness\/}, with maximum stiffness $|R\rt| = 1$ 
expressing time-preserved coherence in the sense of Eq.~(\ref{eq:TCO}). Note 
that the formal employment of subsidiary $(N-1)$-particle states is necessary 
only to provide a reference for the evolution of the true $N$-boson system: 
We do not violate particle number conservation, and hence do not involve 
spontaneous symmetry breaking~\cite{Gardiner97,CastinDum98,Leggett01}.
Moreover, an interesting observation can be made here: If the decisive
relation~(\ref{eq:TCO}) is satisfied, then the definition~(\ref{eq:TDO})
immediately yields
\begin{equation}
	R\rt = \frac{\Phi\rt}{| \Phi\rt |} \; ,
\label{eq:PHS}	
\end{equation}
meaning that the phase of $R\rt$ equals that of $\Phi\rt$. Read in the reverse
direction, this implies that the phase of a macroscopic wave function contains
information on the difference of the evolution of ``neighboring'' $N$-- and 
$(N-1)$-particle states. This is well known in the equilibrium case, when
the phase of the solution to the Gross-Pitaevskii equation is determined
by the chemical potential, {\em i.e.\/}, by the energy required to add one
more particle to the system. The present considerations show that the phase
retains a similar meaning even in case of nonequilibrium, in which a chemical
potential does not exist.

%%%%%%%%%%%%%%%%%%%%%%%%%%%
%% NUMERICAL SIMULATIONS
%%%%%%%%%%%%%%%%%%%%%%%%%%%

\section{Numerical simulations}  
\begin{figure}[t]
\includegraphics[width = 1.0\linewidth]{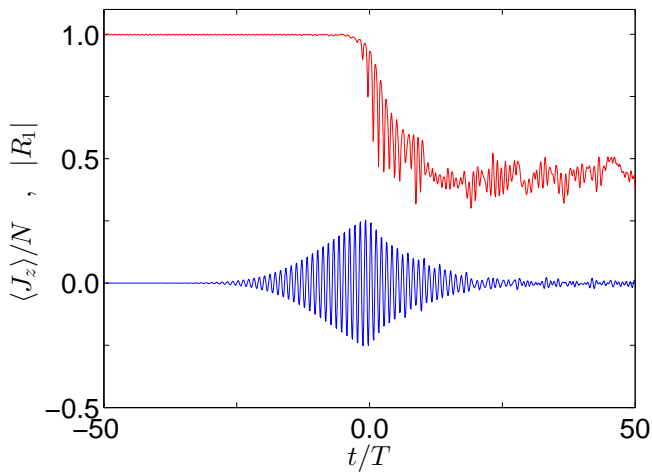}
\caption{{\textbf{Degradation of the order parameter.}} Shown are the 
	 stiffness $|R_1|$ (above) and the scaled population 
	 imbalance~(\ref{eq:INP}) (below) for the driven bosonic 
	 Josephson junction~(\ref{eq:HDR}) with $N = 100$ particles and 
	 scaled interaction strength $N\kappa/\Omega = 2.0$, responding to 
	 a pulse with carrier frequency $\omega/\Omega = 1.6$ and Gaussian 
	 envelope~(\ref{eq:PUL}) with width $\sigma/T = 10$ and maximum 
	 driving strength $\mu_{\rm max}/\Omega = 0.51$. The time scale is 
	 given by the cycle time $T = 2\pi/\omega$. The initial state was the 
	 ground state of the undriven junction~(\ref{eq:HBJ}). Observe that 
	 the macroscopic wave function remains well preserved until the middle 
	 of the pulse, after which the decrease of stiffness signals its 
	 degradation.}
\label{F_2}
\end{figure}

In order to illustrate some consequences of the concepts developed above, 
we utilize the model of a bosonic Josephson junction~\cite{Leggett01,
GatiOberthaler07}, as described by the Hamiltonian     
\begin{equation}
	H_0 = -\frac{\hbar\Omega}{2}\left( \aLd\aR + \aRd\aL\right)
	+ \hbar\kappa\left(\aLd\aLd\aL\aL + \aRd\aRd\aR\aR\right) \; .  
\label{eq:HBJ}
\end{equation}
Here the hopping matrix element between the two sites labeled $1$ and $2$ 
is given by $\hbar\Omega/2$, so that $\hbar\Omega$ is the single-particle 
tunneling splitting, while $2\hbar\kappa$ quantifies the repulsion energy of 
two particles occupying a common site. The bosonic operator $\aj$ annihilates 
a particle at the $j$th site; $\ajd$ is its adjoint creation operator. This 
system is subjected to a time-dependent bias with carrier frequency~$\omega$ 
and envelope $\hbar\mu(t)$, as specified by 
\begin{equation}
	H_1(t) = \hbar\mu(t)\sin(\omega t)\left(\aLd\aL - \aRd\aR \right) \; ;
\end{equation}
the total Hamiltonian then reads
\begin{equation}
	H(t) = H_0 + H_1(t) \; . 
\label{eq:HDR}
\end{equation}
Even with constant amplitude $\mu(t) = \mu_1$ this model captures nontrivial
features of many-body dynamics~\cite{WeissTeichmann08,GertjerenkenHolthaus14};
it is one of the very rare systems which allows one to monitor the emergence 
of a macroscopic wave function numerically, but without further approximations
on the $N$-particle level. For all following simulations we select the ground 
state $| \Psi^{(0)} \rangle_N$ of the time-independent junction~(\ref{eq:HBJ}) 
with scaled interaction strength $N\kappa/\Omega = 2.0$ as the initial state.
Note that this ground state is no $N$-fold occupied single-particle state in 
the sense of Eq.~(\ref{eq:NPC}), because the relatively strong interparticle
interaction leads to sizeable depletion~\cite{GertjerenkenHolthaus15}. 
The required subsidiary states~(\ref{eq:SUB}) then are given by
\begin{equation}
	| \wP_j^{(0)} \rangle_{N-1} = 
	\frac{\aj | \Psi^{(0)} \rangle_N}{\| \aj | \Psi^{(0)} \rangle_N \|}
\end{equation} 
for $j = 1,2$. Moreover, we fix the scaled carrier frequency 
$\omega/\Omega = 1.6$, and consider a Gaussian envelope  
\begin{equation}
	\mu(t) = \mu_{\rm{max}}\exp(-t^2/2\sigma^2)
\label{eq:PUL}
\end{equation}
with width $\sigma/T = 10$, where the time scale is set by $T = 2\pi/\omega$. 
In Fig.~\ref{F_2} we monitor the response of a system with $N = 100$ particles 
to a pulse with maximum driving strength $\mu_{\rm max}/\Omega = 0.51$ by
plotting the scaled population imbalance 
\begin{equation}
	\langle J_z \rangle(t)/N = 
	{_{N}\langle} \Psi(t) | \aLd \aL - \aRd \aR | \Psi(t) \rangle_N / (2N)
\label{eq:INP}
\end{equation}
vs.\ time. We also show the stiffness $|R_1(t)|$; the corresponding quantity
$|R_2(t)|$ obtained for the other site looks practically identical. Although
$N = 100$ is not ``macroscopically large'', one observes that $|R_1|$ stays
close to unity almost until the pulse's middle, and then decreases in an
oscillating manner. Thus, already in this situation there exists a good 
macroscopic wave function during the first half of the pulse, but it degrades 
significantly during the second half.

\begin{figure}[t]
\includegraphics[width = 1.0\linewidth]{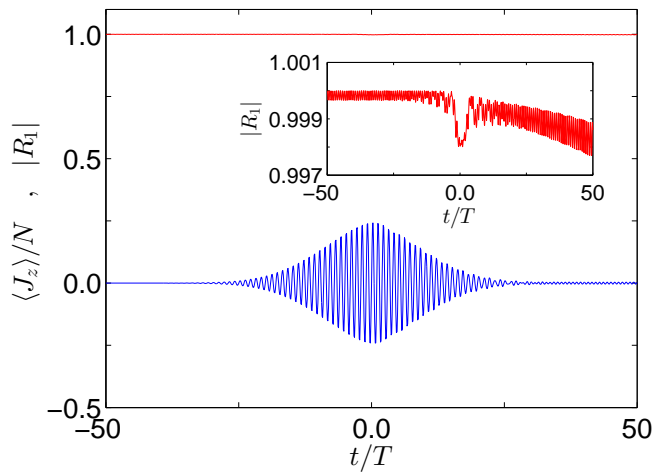}
\caption{{\textbf{Preservation of the order parameter.}} As Fig.~\ref{F_2},
	but with $N = 1000$. Here the macroscopic wave function does hardly
	degrade during the entire pulse. Observe the scale of the insets'
	ordinate!}  
\label{F_3}	
\end{figure}

Increasing the particle number to $N = 1000$, while keeping $N\kappa/\Omega$
and all other parameters constant, we obtain Fig.~\ref{F_3}. This is a truly 
remarkable finding: Although the $N$-particle state undergoes violent changes 
when adjusting itself to the driving force, the stiffness remains close to its 
theoretical maximum during the entire pulse, indicating that one can subject 
a macroscopic wave function to strong forcing almost without reducing its 
order.

\begin{figure}[t]
\includegraphics[width = 1.0\linewidth]{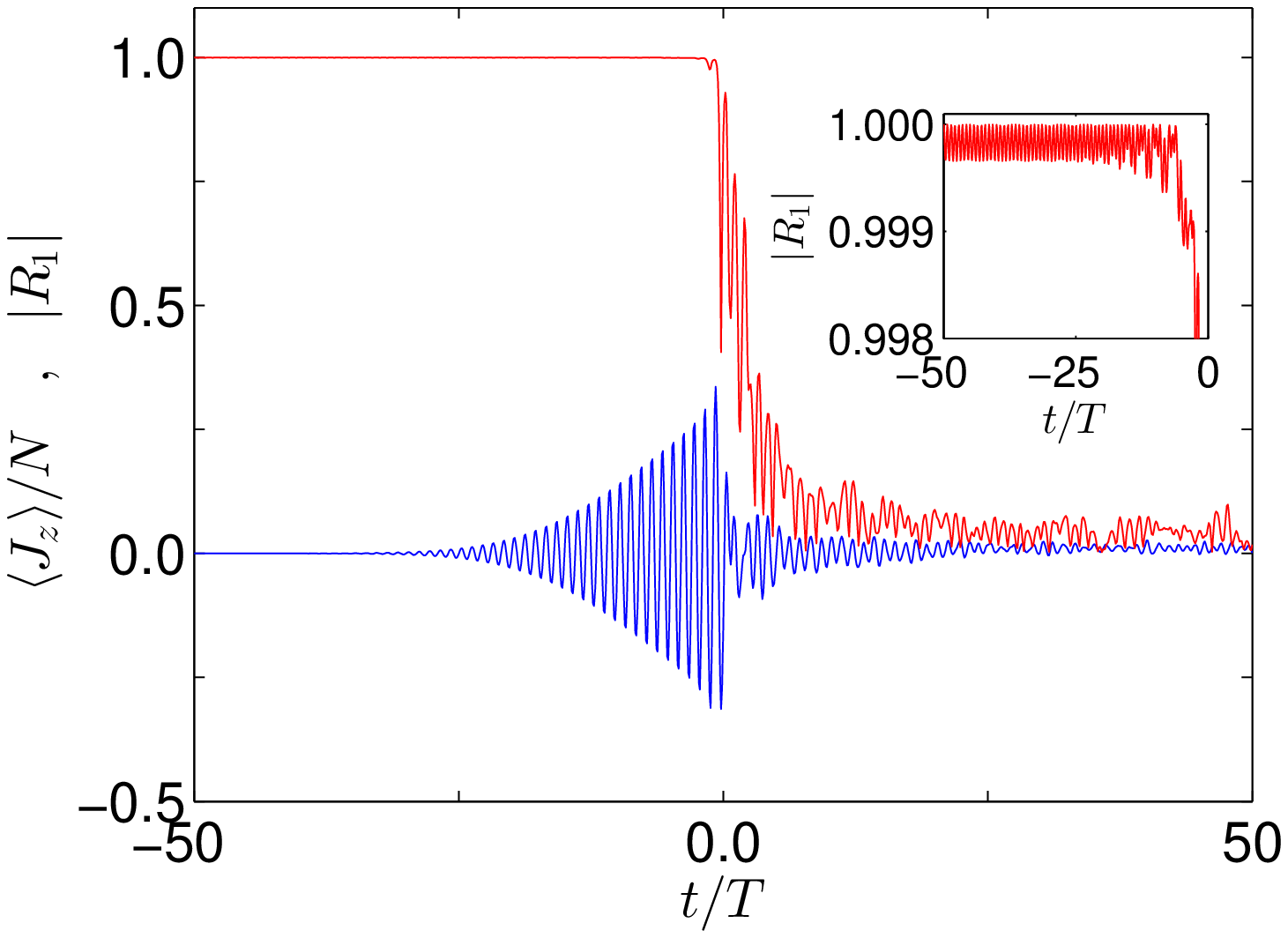}
\includegraphics[width = 1.0\linewidth]{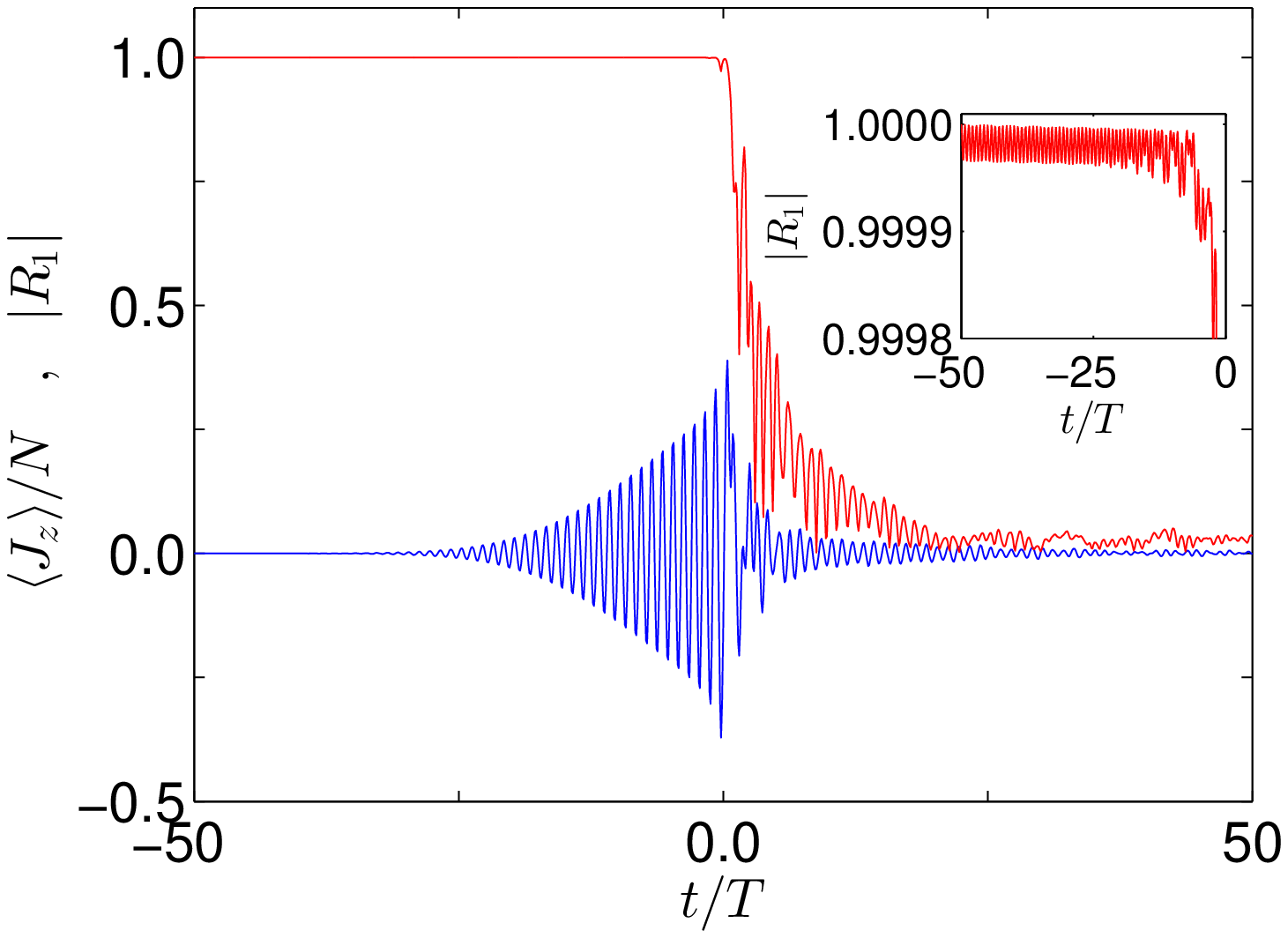}
\caption{{\textbf{Dynamically induced destruction of macroscopic wave 
	functions.}} As Fig.~\ref{F_2}, but with higher driving amplitude
	$\mu_{\rm{max}}/\Omega=0.55$, and $N = 1000$ (upper panel) or 
	$N = 10000$ (lower panel). The macroscopic wave function is not 
	destroyed gradually, but quite suddenly; this destruction cannot be 
	prevented by increasing the particle number.} 
\label{F_4}	
\end{figure}

\begin{figure}[t]
\includegraphics[width = 1.0\linewidth]{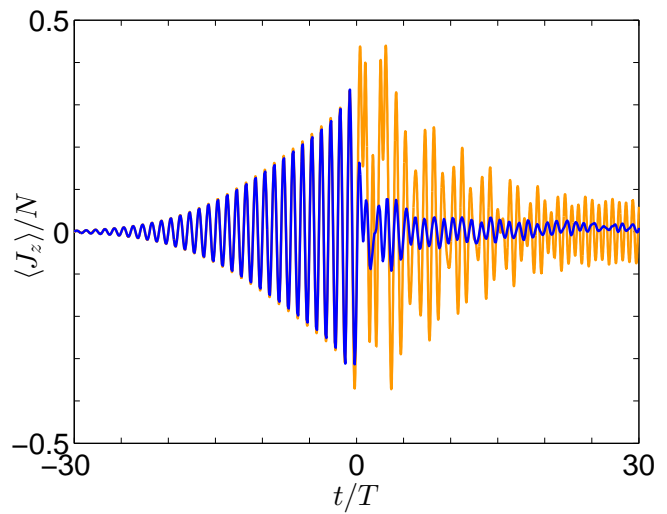}
\caption{{\textbf{Gross-Pitaevskii vs.\ $N$-particle dynamics.}} 
	The $N$-particle population imbalance~(\ref{eq:INP}) for 
	$\mu_{\rm{max}}/\Omega=0.55$ and $N = 1000$, already recorded in the 
	upper panel of Fig.~\ref{F_4}, is compared to the prediction of the 
	Gross-Pitaevskii equation. As long as the stiffness is close to unity, 
	there exists a macroscopic wave function which is perfectly described 
	by the Gross-Pitaevskii equation, so that both curves almost coincide. 
	When the macroscopic wave function is destroyed the solution to the 
	Gross-Pitaevskii equation becomes chaotic, and does no longer predict 
	the $N$-particle dynamics correctly.}
\label{F_5}
\end{figure}

A quite different scenario is depicted in Fig.~\ref{F_4}. Here we have 
increased the driving amplitude to $\mu_{\rm max}/\Omega = 0.55$, and consider
both $N = 1000$ (upper panel) and $N = 10000$ (lower panel). While we observe
excellent stiffness during the first half of the pulse, with $1 - |R_1(t)|$
apparently scaling with $1/N$, the macroscopic wave function is destroyed
suddenly; this sudden destruction {\em cannot\/} be counteracted by an 
increase of~$N$~\cite{GertjerenkenHolthaus15}. We utilize this example also
to illustrate one more feature: As long as it exists, the macroscopic wave 
function should conform to the Gross-Pitaevskii equation. One may still solve 
that equation even beyond the point of destruction of the macroscopic wave 
function, but then the solution no longer captures the actual $N$-particle 
dynamics. This is verified by Fig.~\ref{F_5}, where we superimpose the 
$N$-particle imbalance~(\ref{eq:INP}) for $N = 1000$ to the prediction made by 
the Gross-Pitaevskii equation. As long as there is close-to-perfect stiffness, 
both curves are almost indistinguishable from each other, confirming the 
accuracy of the Gross-Pitaevskii approach under conditions of time-preserved 
coherence. But when the macroscopic wave function is destroyed the 
Gross-Pitaevskii dynamics become chaotic, losing their connection to the 
$N$-particle level.

%%%%%%%%%%%%%%%%%%%%%%%%%%%
%% DISCUSSION
%%%%%%%%%%%%%%%%%%%%%%%%%%%

\section{Discussion}

The observations made in this work have both conceptual and experiment-oriented
consequences. We have addressed the question why the solution to the 
time-dependent nonlinear Gross-Pitaevskii equation can provide a good 
description of a forced condensate only when it behaves in a regular, 
non-chaotic manner: That distinction between order and chaos should have 
a counterpart already on the linear $N$-particle level. As one possible 
characterization of this difference we suggest to monitor the time evolution 
of ``neighboring'' trajectories in Fock space of states consisting of $N$ and 
$N-1$ particles, respectively. With $N$-particle states being orthogonal to 
states consisting of one particle less, the required measure of proximity of 
these states is provided by the projection of the former after annihilation 
of one particle onto the latter. In this way, one can not only give a more 
definite meaning to the sketch by Lifshitz and Pitaevskii on how to construct 
the wave function of the condensate~\cite{LaLifIX}, but one also obtains the 
desired indicator for the quality of this construction: The presence of a 
time-dependent macroscopic wave function necessarily requires that initially 
close state trajectories stay close to each other in the course of time. If 
this condition is satisfied, the unmodified Gross-Pitaevskii equation provides 
an excellent description of the $N$-particle dynamics; if not, the macroscopic 
wave function is destroyed~\cite{GertjerenkenHolthaus15}. 
 
Our matter-of-principle discussion is of little practical help when it comes 
to computing the instability of a driven Bose-Einstein condensate in 
experimentally realistic situations, implying that knowledge of the exact 
$N$-particle state cannot be obtained. In such cases one requires other
approaches, such as the second-order number-conserving self-consistent
treatment developed by Gardiner and Morgan~\cite{GardinerMorgan07}, which 
has been applied to a toroidally trapped, $\delta$-kicked condensate by
Billam {\em et al\/.}~\cite{BillamGardiner12,BillamEtAl13} One then couples
the solution of a generalized Gross-Pitaevskii equation to modified
Bogoliubov-de Gennes equations, assuming that the ratio of noncondensate
to condensate particle numbers be a small parameter. This approach allows
one to assess driven condensate dynamics with experimentally realistic
particle numbers~\cite{GardinerMorgan07,BillamGardiner12,BillamEtAl13}.

Yet, even our idealized model calculations, which are not tied to any
small parameter, do convey messages of practical importance. We have shown that
a driving force does not necessarily destroy a macroscopic wave function when 
it is applied smoothly, in the form of forcing pulses with a sufficiently 
slowly changing envelope. This particular manifestation of the quantum 
adiabatic principle signals green light for systematic quantum engineering 
with macroscopic wave functions. The identification of maximum stiffness, or 
of time-preserved coherence in the sense of Eq.~(\ref{eq:TCO}), as the salient 
feature of a time-dependent macroscopic wave function may guide future 
investigations. Our simulations also illustrate an important fact: The initial 
$N$-particle state considered therein, which is the ground state of the 
model~(\ref{eq:HBJ}), equals a pure condensate state only for vanishing 
interaction, that is, for $N\kappa/\Omega = 0$~\cite{GertjerenkenHolthaus15}, 
whereas we consider strong interparticle interaction, $N\kappa/\Omega = 2.0$. 
Nonetheless, maximum stiffness can still be attained to an amazing degree of 
accuracy, as exemplified in Fig.~\ref{F_3}. Finally, the observation that 
substantial degradation of the underlying order parameter may not occur 
gradually in time, but rather can be connected to certain critical driving 
strengths, is open to experimental verification. Such experiments do not 
necessarily require a driven bosonic Josephson junction, but can also be 
performed in other configurations. For instance, one could subject a 
Bose-Einstein condensate in a strongly anharmonic trap to a smooth forcing 
pulse, and perform a time-of-flight measurement of the condensate fraction 
after the pulse is over. If one repeats this measurement with successively 
stronger pulses, one should observe a sudden disappearance of the condensate 
peak at a certain critical maximum driving amplitude.

\begin{acknowledgments}
	We acknowledge support from the Deutsche Forschungsgemeinschaft (DFG)
	through grant no.\ HO 1771/6-2.	
	The computations were performed on the HPC cluster HERO, located at 
	the University of Oldenburg and funded by the DFG through its Major 
	Research Instrumentation Programme (INST 184/108-1 FUGG), and by the 
	Ministry of Science and Culture (MWK) of the Lower Saxony State.
\end{acknowledgments}

\end{document}